 \documentclass[preprint,3p,times,twocolumn]{elsarticle}

\usepackage{graphicx}      % include this line if your document contains figures
\usepackage{natbib}        % required for bibliography
\usepackage{amsthm}
\usepackage{amsmath}
\usepackage{amsfonts}
\usepackage{comment}
\usepackage{cancel}
\usepackage{hyperref}

\hypersetup{
dvips,
backref=true, %permet d'ajouter des liens dans...
pagebackref=true,%...les bibliographies
hyperindex=true, %ajoute des liens dans les index.
colorlinks=true, %colorise les liens
breaklinks=true, %permet le retour \`a la ligne dans les liens trop longs
urlcolor= PresqueNoir, %couleur des hyperliens
linkcolor=PresqueNoir, %couleur des liens internes
citecolor=PresqueNoir,
bookmarks=true, %cr\'e\'e des signets pour Acrobat
bookmarksopen=true}

\newtheorem{rem}{Remark}[section]

\newcounter{moncompteur}
\newcounter{compp}
\newcounter{comppp}
 \newcounter{cpfig} 
\newcommand{\CPTFIG}{\stepcounter{cpfig} Fig.\thecpfig~}
\newcommand{\CPTID}{ Fig. \themoncompteur  \ }
\newcommand\CPTIDD[1]{\setcounter{compp}{\thecpfig}\addtocounter{compp}{#1}Fig.\thecompp}
\newcommand{\nablabf}{\boldsymbol{\nabla}}
\newenvironment{centered}
 {\begin{list}{}{\leftmargin=0cm
                  \itemsep=0ex
                  \parsep=0pt
                  \topsep=0pt
                  \parskip=0pt
                     \footnotesize}
\centering\item\relax}
  {\end{list}}

\journal{Computer Physics Communications}

\begin{document}
\begin{frontmatter}

\title{Efficient periodic  band diagram computation using a finite element method, Arnoldi eigensolver and sparse linear system solver.}
\author[ON,UPS]{ Romain Garnier}
\ead{romain.garnier@inria.fr}
\author[ON]{Andr\'e Barka}
\ead{andre.barka@onera.fr}
\author[UPS,CNRS]{ Olivier Pascal}
\ead{ olivier.pascal@laplace.univ-tlse.fr}
\address[ON]{ONERA The French Aerospace Lab - 2 av. Edouard Belin BP 4025 - 31055 Toulouse }

\address[UPS]{Universit\'e de Toulouse ; UPS, INPT ; LAPLACE (Laboratoire Plasma et Conversion d \'Energie) ; \\
118 route de Narbonne, F-31062 Toulouse cedex 9, France. }
\address[CNRS]{CNRS ; LAPLACE ; F-31062 Toulouse, France.}

\begin{abstract}                % Abstract of not more than 250 words.
We present here a Finite Element Method devoted to the simulation of 3D periodic structures of arbitrary geometry. The numerical method based on ARPACK and PARDISO libraries, is discussed with the aim of extracting the eigenmodes of  periodical structures and thus establishing their frequency band gaps. Simulation parameters and the computational optimization are the focus. Resolution will be used to characterize EBG (Electromagnetic Band Gap) structures, such as  plasma rods  and metallic cubes. 
\end{abstract}

\begin{keyword}
%Five to ten keywords, preferably chosen from the IFAC keyword list:
 Finite element method, Eigenvalue problems, Periodic structures, Electromagnetic Band Gap, Sparse matrices.
\end{keyword}

\end{frontmatter}
%===============================================================================

\section{Introduction}
%\begin{minipage}{0.65\linewidth}
The  classical plane wave method \cite{Cristophoto} allows  the band diagram of EBG (electromagnetic band gap)
structures, like photonic crystal, to be established. These materials are assumed to be periodic, non-dispersive and
dielectric ($\epsilon >0$). However modifications of the plane wave method allow dispersive media \cite{Disper1,Disper2,Disper3} to be dealt with. Especially, the plane wave method can also be used to characterize surface mode solutions \cite{FloquetBloch,PeriodicBook}. However this method, often used in these cases, may be highly difficult to develop when the periodic patterns of the material have an arbitrary shape and permittivity. In this case the Finite
Element Method, together with an eigenmode solver, is very efficient. One can, for instance, note the
capability of the techniques based on the mesh refinement used by the ANSOFT commercial software HFFS to process it. However, by using this software you cannot control all of the eigenmode resolution parameters \cite{HFFSHELP1,HFFSHELP2}. Due to the existence of the spurious modes \cite{spurious}, the convergence can become very questionable and there is a lack of interpretation and investigation tools to overcome this point\cite{EUCAP} and we want to give a detailed explanation about the way to configure the parameters involved in the calculation. 
The purpose of this study is to propose an effective approach to establish the band-diagram of complex periodical structures.
In Section \ref{Scbas} we will explain the theoretical approach of the method, in Section \ref{Approach}  we will explain the way to optimize the Arnoldi algorithm in order to sketch the band-diagram and in Section \ref{RESULT} we will present selected results that illustrate the method.

%\end{minipage}
%\begin{minipage}{0.35\linewidth}
%\centering
%\footnotesize
\begin{centered}
\includegraphics[width=0.6\linewidth]{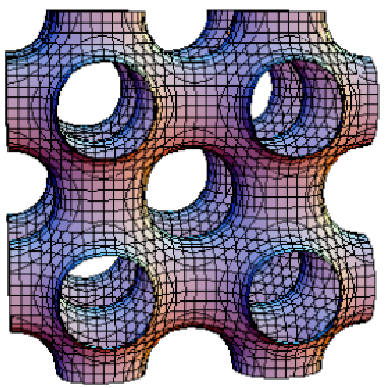}\\
\CPTFIG Example of a 3d periodic structure with complex shape\cite{Marianne}.
\end{centered}

\section{Scientific basis}\label{Scbas}
Materials are assumed to be non dispersive in this work, that is to say the permittivity is assumed to be non frequency dependent.
We are proposing a code that uses a finite element method
to solve eigenmode problems for periodic structures. It will
also allow convergence parameters  to be controlled  at each step
of the resolution.
For all test function $\boldsymbol{\varphi}\in H(curl,\Omega)$ the code is based on the weak
formulation \cite{FEM} :
\begin{small}
\begin{equation}\label{WEKFORM}\int_{\Omega} \frac{\nabla \times \mathbf{E}}{\mu_r}.\nabla\times\boldsymbol{ \bar{\varphi}}=k_0^2\int_{\Omega}\epsilon_r\mathbf{E}.\boldsymbol{\bar{\varphi}}-\boxed{\int_{\partial\Omega}\big(\boldsymbol{\nu}\times(\frac{\nabla \times \mathbf{E}}{\mu_r})\big).\boldsymbol{\bar{\varphi}}}\end{equation}
\end{small}
Then we can define the bi-linear forms~:
\[a(\mathbf{E},\boldsymbol{\varphi})=\int_{\Omega} \frac{\nabla \times \mathbf{E}}{\mu_r}.\nabla\times\boldsymbol{ \bar{\varphi}}+\int_{\partial\Omega}\big(\boldsymbol{\nu}\times(\frac{\nabla \times \mathbf{E}}{\mu_r})\big).\boldsymbol{\bar{\varphi}},\]
\[ m(\mathbf{E},\boldsymbol{\bar{\varphi}})=\int_{\Omega}\epsilon_r\mathbf{E}.\boldsymbol{\bar{\varphi}}\]
Where $\boldsymbol{\bar{\varphi}}$ denotes the complex conjugate of $\boldsymbol{\varphi}$.

The unknowns are the electric $\mathbf{E}$ field and the $k_0$ wave number. $\boldsymbol{\nu}$ is the outside normal of the interface
$\partial \Omega$. The relative permittivity and permeability are denoted by respectively $\epsilon_r$  and $\mu_r$. We further assume that both relative permeability and relative permittivity are $(3\times 3)$ positive definite Hermitian tensors.
After we mesh the domain we use the Nedelec zero order edge functions\cite{Nedelec} on each edge $[i,j]$~:
\[\mathbf{W}_{ij}=(\lambda_i\mathbf{\nabla}\lambda_j-\lambda_j\mathbf{\nabla}\lambda_i)*l_{ij} \]
Where $\lambda_i$ are the barycentric coordinates of the point $i$, and $l_{ij}$ is the length of the edge $[i,j]$.
We call $\mathcal{V}_h$ the subspace generated by these basis functions. Then an approximation of the $\mathbf{E}$ field can be expressed as a linear combination of these functions~:
\begin{equation}\label{Fbase} \mathbf{E}\simeq \mathbf{\tilde E}=\sum_{i\neq j} e_{[ij]}\mathbf{W}_{ij} \mathrm{ \ where \ } e_{[ij]}=\int_{[ij]}\mathbf{\tilde E}.\end{equation}
 By calling \textbf{X} the coordinates of $\mathbf{\tilde E}$ in  $\mathcal{V}_h$  and by applying the Galerkin method, we obtain the linear system~:
\begin{equation}\label{lineaire}\begin{array}{c} A\mathbf{X}=k_0^2M\mathbf{X}, \mathrm{ \ where \ }\\
 A_{pk}=a(\mathbf{W}_p,\mathbf{W}_k), \ M_{pk}=m(\mathbf{W}_p,\mathbf{W}_k), \\
 p,k\in [1\ldots DF], \ \mathbf{X}=(e_1, \ldots e_{DL}) \end{array} \end{equation}
\begin{rem} \label{SPD} If we erase the boxed term in \eqref{WEKFORM} then the $A$ and $M$ matrices will be positive semi-definite by construction. \end{rem}

$DF$ are the degrees of freedom of the $\mathbf{\tilde E}$ field, it depends on the boundary conditions. For example, when we impose perfectly electric conducting boundary conditions the integration term on $\partial\Omega$ disappears and all of the values $\{e_{p}, \ [p] \subset \partial\Omega\}$ 
are equal to zero, the number of unknowns are $\{e_p, \ [p] \subset  \ \stackrel{\circ}{\Omega}\}$.  
The purpose is then to compute the eigenmodes first, for metallic cavities (P.E.C boundary conditions) and next for periodic structures by adding PBC (Periodic Boundary Conditions).\newline
After finding solutions for cavities, we have modified matrices initially built to solve problems into periodic structures, by applying the Floquet conditions \cite{Cristophoto,Periodique3D,Pyati} on $F_m$(master) and $F_s$(slave) translated by the vector $\mathbf{T}$~:

\begin{equation}\label{Floquet}\mathbf{E}_{F_s}=e^{-j\phi}\mathbf{E}_{F_m}, \ \ \phi=\mathbf{k}_i.\mathbf{T}\end{equation}

where $\mathbf{k}_i$ is the incident wave vector, which is an initial condition of the problem, $\mathbf{E}_{F_s}$
is the field on the slave face and  $\mathbf{E}_{F_m}$ the field on the master face.
By definition the band-diagrams are the frequency solutions of Maxwell equations for each given value of $\phi\in[0,\pi].$ As the structure is periodic we solve the Maxwell equation on the unit cell of the periodic structure. Let us take an example with a simple bi-periodical structure~:

\begin{centered}
\includegraphics[width=0.7\linewidth, height=0.6\linewidth]{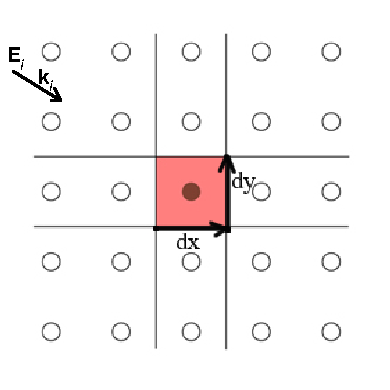}\\
\CPTFIG In red the basic cell of a 2-dimensional periodic structure.\\
$\mathbf{T}_1=(dx,0), \ \mathbf{T}_2=(0,dy), \ \mathbf{k}_i=(k_{ix},k_{iy}).$
\end{centered}
All the points $\tilde{\mathbf{P}}$ in the infinite periodic domain have one corresponding point $\mathbf{P}$ in the red domain and if we know the values of the field  $\underline{\mathbf{E}}$ in this domain, all the solution in the infinite periodic domain can be reproduced using Floquet conditions~:
\[\exists (m,n) \in \mathbb{Z}^2, \ \tilde{\mathbf{P}}=\mathbf{P}+m\times \mathbf{T}_1+ n\times \mathbf{T}_2\Rightarrow\]
\[\mathbf{E}(\tilde{\mathbf{P}})=\underline{\mathbf{E}}(\mathbf{P})e^{-j\big(m\times \mathbf{T}_1+n\times \mathbf{T}_2\big).\mathbf{k}_i}. \]
Moreover, the field can be written as a pseudo periodical series \cite{Cristophoto}~:
  \[\mathbf{E}(x,y)=\sum_{n=-\infty}^{+\infty}\sum_{m=-\infty}^{+\infty} \mathbf{a}_{nm} e^{j\big((\frac{2\pi n }{d_x}-k_{ix})x+(\frac{2\pi m}{d_y}-k_{iy})y\big)}\]
\[ \mathbf{a}_{nm}=\frac{1}{d_xd_y}\int_0^{d_x}\int_0^{d_y} \underline{\mathbf{E}}(x,y)e^{-j\big((\frac{2\pi n }{d_x}-k_{ix})x+(\frac{2\pi m}{d_y}-k_{iy})y\big)}dxdy \]

\subsection{Consequences on the initial system} 
In this section, we want to show the transformation of the system \eqref{lineaire} induced by the Floquet conditions \eqref{Floquet}.
As we have already seen in equation \eqref{Floquet} fields of opposite interfaces are linked by phase shifts and thus, the unknown of the system \eqref{lineaire} can be linked as follow~:
\begin{equation}\label{Reduction}e_{M_k}=e^{-j\phi}e_{S_k}, \ \phi=\mathbf{k}_i.\mathbf{T}, \ k \in [1\ldots N_{M}]\end{equation}
 $e_{M_k}, \  k \in [1\ldots N_{M}]$ are the unknowns located on $F_m$(master) and  $e_{S_k}$ the corresponding unknowns on $F_s$(slave).

The equation \eqref{Reduction} shows that the periodic boundary conditions lead to a reduction in the number of unknowns. As a consequence the approximation subspace will change \cite{feriera} and we will define the new basis functions~:
\[ \begin{array}{c}
\tilde{\mathbf{W}}_{I_i}=\mathbf{W}_{I_i}, \tilde{\mathbf{W}}_{M_k}=\mathbf{W}_{M_k}+e^{-j\phi}\mathbf{W}_{S_k} \\ 
 i \in[1\ldots N_I], \ \ \ k \in [1\ldots N_{M}]. \end{array}\] 

Where $I_i$ denotes the indices of the internal edges, $M_k$ denotes the indices of the edges which are located on the master faces and $S_k$ denotes the indices of the edges which are located on the corresponding slave faces. 
As a consequence, the variational problem can be written as follows~:
\begin{footnotesize}
\[\sum_{i=1}^{N_I}e_{I_i}*a(\mathbf{W}_{I_i},\tilde{\mathbf{W}}_{q})+\sum_{k=1}^{N_{M}}e_{M_k}*\big(a(\mathbf{W}_{M_k},\tilde{\mathbf{W}}_q)+ e^{-j\phi}*a(\mathbf{W}_{S_k},\tilde{\mathbf{W}}_q)\big)=\]
\[k_0^2\big[\sum_{i=1}^{N_I}e_{I_i} *m(\mathbf{W}_{I_i},\tilde{\mathbf{W}}_q)+ \sum_{k=1}^{N_{M}}e_{M_k} *\big(m(\mathbf{W}_{M_k},\tilde{\mathbf{W}}_q)+ e^{-j\phi}*m(\mathbf{W}_{S_k},\tilde{\mathbf{W}}_q)\big)\big] \]
\[(q=I_i, \ i\in[1\ldots N_I])\cup(q=M_k, \ k\in [1\ldots N_{M}])\]
\end{footnotesize}
 \begin{rem}
The terms $a(\mathbf{W}_{p}, \tilde{\mathbf{W}}_{q}) $ will be non-zero terms if the edges $[p]$ and $[q]$ are in the same tetrahedron.
\end{rem}
\begin{rem}
The approximation space is equipped with a Hermitian structure~:
\[a(\mathbf{W}_{p},\tilde{\mathbf{W}}_{M_k})=a(\mathbf{W}_{p},\mathbf{W}_{M_k})+e^{j\phi}a(\mathbf{W}_{p},\mathbf{W}_{S_k})\]
\end{rem}

More precisely, let us consider a bi-periodical structure, the unknowns are linked by the phase shift relations developed below.

\begin{centered}
\includegraphics[width=1\linewidth]{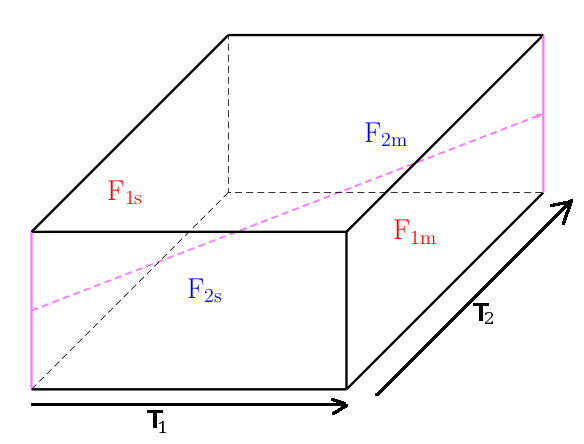}\\
 % \begin{pspicture}(-2.5,-1.2)(5,3.6)
%\psframe[fillstyle=solid,%
%fillcolor=lightgray](0.5,1.5)(4.5,1.5)(4.5,3.5)(0.5,3.5)
%\psline[linewidth=0.04cm,linecolor=Purple](4.5,1.5)(4.5,3.5)
%\psline[linewidth=0.02cm,linecolor=Purple,linestyle=dashed]{->}(-2,0)(4.5,2.5)
%\psline[linewidth=0.04cm](-2,-1)(2,-1)(-2,-1)(2,-1)
%\psline[linewidth=0.04cm](2,1)(-2,1)
%\pspolygon[linewidth=0.02cm,linestyle=dashed](0.5,1.5)(4.5,1.5)(4.5,3.5)(0.5,3.5)
%\psline[linewidth=0.04cm]
%\psline[linewidth=0.04cm,linecolor=Purple](-2,-1)(-2,1)
%\psline[linewidth=0.04cm](2,-1)(2,1)
%\psline[linewidth=0.02cm,linestyle=dashed](-2,-1)(0.5,1.5)
%\psline[linewidth=0.04cm](2,-1)(4.5,1.5)
%\psline[linewidth=0.04cm](2,1)(4.5,3.5)
%\psline[linewidth=0.04cm](-2,1)(0.5,3.5)
%\psline[linewidth=0.04cm](4.5,3.5)(0.5,3.5)
 %\uput[180](-0.5,1.5){\color{red}$F_{1s}$} 
%\uput[180](0,3){$\tilde{d}$} 
%  \uput[0](-2.7,0){ dy}
 %\uput[0](2.5,1){\color{red}$F_{1m}$} 
 %\uput[90](2.5,2){\color{blue}$F_{2m}$}
%   \uput[90](1,1.7){\color{green}$F_{2m}$}
    
 %\uput[-90](0.2,0.5){\color{blue}$F_{2s}$} 
%  \uput[-90](1.7,0.8){\color{green}$F_{2s}$} 
%      \uput[-90](0.5,-1){dx}

%\end{pspicture}
\CPTFIG Phase shift relationships.
\end{centered}

The unknowns $e_{1s}$(respectively $e_{2s}$) on the edges on the slave face  $F_{1S}$ (respectively $F_{2S}$) are equal to $e^{-j\phi_1}*e_{1m}$ (respectively $e^{-j\phi_2}*e_{2m}$), where $e_{1m}$ (respectively $e_{2m}$) are the unknowns of the field on the face $F_{1m}$ (respectively $F_{2m}$). The phase shifts $\phi_1$ and $\phi_2$ are  respectively equal to  $\mathbf{k}_i.\mathbf{T}_1$ and  $\mathbf{k}_i.\mathbf{T}_2$. We see that the edges at the intersection of two slave faces $F_{1s}$ and $F_{2s}$ can be correlated with the edges of the intersection of the two master sides $F_{1m}$ and $F_{2m}$ and the phase shift will be equal to $-(\phi_1+\phi_2)$ \cite{Pyati}.

The transformed matrices $\tilde{A}$ and $\tilde{M}$ by the above method are Hermitian semi-positive definite. Indeed, the above transformations are Hermitian and a matrix $R$ exists such that \cite{TransfoPer}~:
 \[\tilde{A}=RA\bar{R}^T, \ \ \tilde{M}=RM\bar{R}^T\]
 Since the matrices $A$ and $M$ are positive semi-definite (see remark \ref{SPD}) the new build matrices $\tilde{A}$ and $\tilde{M}$ will also remain positive semi-definite.  
 If we denote by 
 \[\begin{gathered} INT=\{I_i,\ldots I_{N_I}, \ \mathbf{edge}[I_i] \subset \ \stackrel{\circ}{\Omega}\}, \\
  \mathcal{F}_{1m}=\{M_1\ldots M_k, \ \mathbf{edge}[M_k] \subset  F_{1m}\} \ldots , \end{gathered}\]
  then $R$ can be written as the bloc matrix~:
 
 \begin{scriptsize}
 \[\begin{matrix}\begin{matrix}  \qquad \qquad\qquad INT  & \mathcal{F}_{1s}  & \mathcal{F}_{1s}\cap \mathcal{F}_{2s}\; \; & \mathcal{F}_{2s}  & \mathcal{F}_{1m} \; \;  & \mathcal{F}_{2m}  &  \mathcal{F}_{1m}\cap \mathcal{F}_{2m}  \end{matrix} \\
 \\
 \begin{matrix} INT \\ \\ \mathcal{F}_{1m}\\ \\ \mathcal{F}_{2m}\\ \\ \mathcal{F}_{1m}\cap \mathcal{F}_{2m} \end{matrix} \begin{bmatrix}   I     & 0  & \ldots & \ldots    &  & &  0 &  &  & 0      \\  & & & & & & & & & \\  0 &      e^{j\phi_1} I & 0 &  0   & I & & 0 & & & 0    \\ & & & & & & & &  & \\ 0  & 0  & e^{j(\phi_1+\phi_2)}I & 0   & 0 &  & I & &  & 0  \\  & & & & & & & & &\\   0 &  \ldots & 0 & e^{j\phi_2}I   & \ldots & & 0   & & & I \end{bmatrix} 
\end{matrix}
\]
 \end{scriptsize}

To do this transformation we must construct an appropriate grid~:

\begin{centered}
\includegraphics[width=1\linewidth]{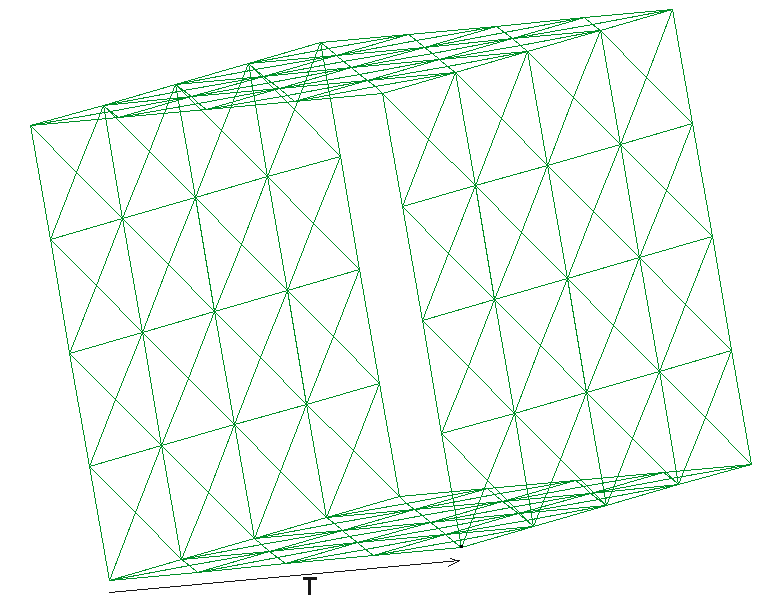}\\
\CPTFIG Master slave grid.\\
\end{centered}

\begin{rem}\label{PerInter}
The mesh is structured so that for each member of the master face there is one member of the slave face translated by the constant vector $\mathbf{T}$. Then, if we consider a triangle $T_m$ on the master face and  the corresponding triangle $T_s$
 on the slave face ($T_m = T_s + \mathbf{T}$), and adding the fact that the tangential component of the basis functions is continuous we have~:
 \begin{small}
 \[ \int_{T_m}\big(\boldsymbol{\nu}_m\times(\frac{\nabla \times \mathbf{W}_{qm}}{\mu_r})\big).\mathbf{W}_{pm}=\textcolor{red}{-}\int_{T_s}\big(\boldsymbol{\nu}_s\times(\frac{\nabla \times \mathbf{W}_{qs}}{\mu_r})\big).\mathbf{W}_{ps} \]
 \end{small}
Therefore, the integrals on the master-slave triangles will vanish in pairs, during the assembly of matrix $A$. For the same reason all of the integrals of the triangles terms will also vanish inside the volume, because the triangles are located at the interface of two tetrahedrons.
\end{rem}
\begin{rem}
In comparison with the plane wave method, there is no need to change the permittivity  value into the perfect metallic media as has be done in\cite{Disper1}.
 Indeed the metallic part will be treated with a PEC condition there is no need to mesh this part.
\end{rem}

\section{Eigen solver strategy}\label{Approach}
\subsection{Storage of the matrices}\label{Storage}
Throughout the rest of the presentation, we will consider that the finite element matrices are constructed using periodic boundary conditions and we will denote these by $A$ and $M$. 
We can also combine PBC and PEC conditions, but we are not dealing with these possibilities in the examples that we choose to treat. 
By construction $A$ and $M$ are sparse, and we must find an
effective way first to store and assemble the matrices. To do this we first have to know the graph of the matrices~:

%\begin{minipage}{0.5\linewidth}
\begin{centered}
\includegraphics[width=1\linewidth]{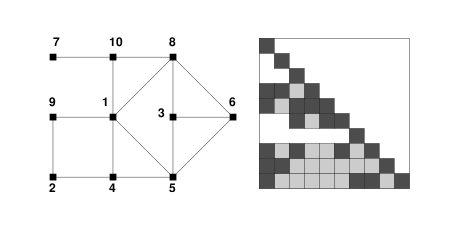}\newline
\CPTFIG Graph associated to the matrix.\newline
\end{centered}

The indices (line $i$, column $j$) of the  non-zero coefficients are those whose edges $(i,j)$ are in a same tetrahedra. Thus by knowing the structure of the mesh we can construct the matrix graph. Thanks to this graph, we can construct the first two arrays $IA$ and $JA$ of the compressed sparse row format \cite{Matlabis} and then assemble the non-zero coefficient stored into an array $\overleftrightarrow{A}$. The dimension of $IA$ is equal to the total number of unknowns plus one while the dimensions of $JA$ and $\overleftrightarrow{A}$ are equal to the total number of the non-zero coefficients.
$JA$ contains the index of the column of each non-zero coefficient while $IA(l)$ contains the position in $\overleftrightarrow{A}$ of the first non-zero coefficient in line $l$ of A. Let us consider an example~:
\[A=
\begin{bmatrix}
1 & 2 & 0 & 0 \\
 0 & 3 & 9 & 0 \\
 0 & 1 & 4 & 0 \\
  0 & 0 & 1 & 2 \\
\end{bmatrix}
\]
is a four-by-four matrix with six non-zero elements, thus
\[
\begin{array}{l}
\overleftrightarrow{A}  = [  \ 1 \ 2 \ 3 \ 9 \ 1 \ 4 \ 1 \ 2 ]\\ 
IA = [ \ 1 \ 3 \ 5  \ 7 \ 9 ] \\    
JA = [ \ 1 \ 2 \ 2 \ 3 \ 2 \ 3 \ 3 \ 4 ].\\
\end{array}
\]
\subsection{Writing the eigenvalue problem}
If the $A$ matrix is invertible, the linear system can be written as an eigenvalue problem~:
\begin{equation}\label{eigenvalue}A\mathbf{X}=k_0^2 M\mathbf{X}\Rightarrow A^{-1}M\mathbf{X}=\frac{1}{k_0^2}\mathbf{X}.\end{equation}
\begin{rem}
Usually the central processing unit CPU time for
the diagonalization of the matrix $A^{-1}M$ is proportional to the number of operations needed for the LU factorization of the A matrix which, is proportional to the cube of its dimension $N$ and hence to the cube of the number of the plane waves if we apply the plane wave method. However, in our case the matrices are sparse and the number of unknowns are defined by the number of edges in the mesh.
Thus, by comparison to the plane wave method the number of  operations and memory used can drastically decrease. Indeed computer storage requirements and the computational complexity of simple array operations are proportional to the number of non-zero elements $nnz(A)$\cite{Matlab}. Looking at the LU factorization for sparse matrices which uses a zero order reduction algorithm \cite{MDRA}, we can deduce that the number of operations for the diagonalization of a sparse matrix should  be proportional to $N\times nnz(L+U)\simeq N\times nnz(A) $.
\end{rem}

 The algorithm we choose to use will give the largest eigenvalues and these eigenvalue will be equal to $\frac{1}{k_0^2}$,
 so we will get the first few smallest values of $k_0$. However, the transformation shown in equation \eqref{eigenvalue} is not suitable because the matrix $A$ is not invertible. Indeed, thanks to the discrete Helmholtz decomposition \cite{Helmholtz}, there exists $\mathbf{U} \in (H^1(\Omega))^3$ derived from a vector potential $\mathbf{V} \in H(curl,\Omega)$ and a scalar potential $\phi \in H^1(\Omega)$ such as~:
\[\mathbf{\tilde E}=\mathbf{U}+\nablabf \phi , \ \mathrm{ such \ as \ } \nablabf . \mathbf{U}=0 \mathrm{ \ and \ } \mathbf{U}=\nablabf\times \mathbf{V}.\]
The kernel of $A$ then contains the fields $\mathbf{\tilde E}$ (defined equation \eqref{Fbase}) which have a zero rotational component (i.e $\mathbf{U} \equiv\mathbf{0}$). More precisely, the solutions of the equation \eqref{lineaire} may be classified into 3 groups \cite{spurious}~:
 \begin{equation}\label{3cas}\left\{
          \begin{array}{l}
\mathrm{Group \ 1 \ \ }k_0^2\neq 0 \mathrm{ \ and \ } \boldsymbol{\nabla} . [\epsilon_r]\mathbf{E}=0  \\  
\mathrm{Group \ 2 \ \ }k_0^2= 0 \mathrm{ \ and \ } \boldsymbol{\nabla} . [\epsilon_r]\mathbf{E}=0 \\
  \mathrm{Group \ 3 \ \ }k_0^2= 0 \mathrm{ \ and \ } \boldsymbol{\nabla} . [\epsilon_r]\mathbf{E}\neq 0 
        \end{array}\right.
\end{equation}
 
 The solutions included in Group 3 are called spurious modes and we want to avoid them. These solutions can be written as a potential-like function  and they have a zero curl. 
When we solve cavity problems, the spurious modes can be eliminated by adding a zero divergence constraint into the Lanczos algorithm \cite{JINFALEE}. When we deal with periodical boundary conditions, the technique used in \cite{JINFALEE} is no longer feasible with zero order Nedelec basis functions because of the constraint on the meshes \cite{PerGrad}. 
%However  If we call by $G$ %the discrete gradient \cite{JINFALEE} constructed by choosing a reference potential on one node of %the slave faces. Once the solutions are constructed, among zero solutions those who verify~:
%\[G^T M\mathbf{X}=0\]
%Have got a null divergence and belong to Group 2. 
In our problems, we will eliminate all of the zero solutions, that is to say, we will keep only the solutions included in Group 1.
As the kernel of the matrix $A$ is not equal to zero, the initial eigenvalue problem \eqref{eigenvalue} must be transformed. The transformation we have decided to apply is the shift and invert transformation, which consist in computing the eigenvalues of the operator $(A-\sigma M)^{-1} M$. If we denote by $\mu$ one eigenvalue of the operator $(A-\sigma M)^{-1}M$, we have~:

\begin{equation}\label{defshift} (A-\sigma M)^{-1} M\mathbf{X}=\mu \mathbf{X} \ \ \mathrm{with} \ \ \mu=\dfrac{1}{k_0^2-\sigma}\end{equation}

Then, if we seek  solutions for the problem $A\mathbf{X}=k_0^2 M\mathbf{X}$ we must look for the greatest values of the problem~:
$OP*\mathbf{X}=\mu \mathbf{X}$ and the values obtained will be close to the \textit{shift} $\sigma$, given $OP=(A-\sigma M)^{-1} M$ 

To do this, we use an improved Arnoldi algorithm proposed by the ARPACK library \cite{ARPACK} on operator $OP$. Arnoldi algorithm consists in orthonormalizing the Krilov space $(\operatorname{span} \, \{ \mathbf{v}_0, OP\mathbf{v}_0, OP^2\mathbf{v}_0, \ldots, OP^{p-1}\mathbf{v}_0 \})$ and is as follows~:
\[\begin{array}{l|l}
\mathbf{w} = OP * \mathbf{v}_j & OP\mathbf{v}_j =\displaystyle\sum_{i=1}^{j+1}h_{ij}\mathbf{v}_i   \\
\mathrm{for \ } i = 1 \mathrm{ \ to \ } j \mathrm{ \ do} &    \\
h_{ij} = (\mathbf{w}.\mathbf{v}_i)  & \mathbf{v}_{j+1} = \frac{\mathbf{w}}{\vert\vert\mathbf{w}\vert\vert} \\
\mathbf{w} =\mathbf{w}-h_{ij}\mathbf{v}_i &     \\
& \\
\mathrm{end \ do} &  h_{j+1,j}=\vert\vert\mathbf{w}\vert\vert 
\end{array}
\]
At each step we have to compute the value of $OP\mathbf{v}_j$.
In other words we must solve the linear system~:
\begin{equation}\label{Systm} (A-\sigma M)\mathbf{X}= M \mathbf{v}_j \end{equation}
 First we have to compute the sparse matrix-vector product $M\mathbf{v}_j$ which will be stored in the vector $Y$ (the notations are the same as in section \ref{Storage})~:
\[
\begin{array}{l}
        Y=0\\
      \mathrm{for \ } i = 1 \mathrm{ \ to \ } n \mathrm{ \ do}\\
      \mathrm{for \ } j = IM(i) \mathrm{ \ to \ } IM(i+1)-1 \mathrm{ \ do}\\
        Y(i)=Y(i)+\overleftrightarrow{M}(j)*\mathbf{X}(JM(j))\\
        END \ DO\\
        END \ DO\\
\end{array}
\]
Here we can see that the cost of operation to perform a matrix vector product is very cheap. 
Then we have to solve the linear system \eqref{Systm}.
  To do this we choose to use a direct sparse solver PARDISO \cite{PARDISO} which uses zero order reducing algorithms to first perform the LU factorization to the matrix $(A-\sigma M)$ and then solve the system. 
As we have triangular matrices (lower and upper) the system can be solved  at each step directly using forward and backward resolution.
\begin{rem}\label{Structure}
In general the structure of the graph of sparse matrices influence the computational time of the LU factorization performed by zero order reducing algorithms. Even if we can optimize these types of algorithms based on profiles of matrices, the graph structure of the matrix $A$ will influence the computational time of solving the system \eqref{Systm}. As we have already noticed in section \ref{Storage}, the graph structure is directly related to the structure of the mesh. We can therefore conclude that the structure of the mesh affects the computation time of the resolution of the system \eqref{Systm}.
\end{rem}

\subsection{Properties of the construction}
After the construction of the first Arnoldi vectors, we have the following properties~:

\[OP\mathbf{v}_j =\sum_{i=1}^{j+1}h_{ij}\mathbf{v}_i.\]

If denote by $V_{\bar{p}}=( \mathbf{v}_1\ldots  \mathbf{v}_{\bar{p}})$,we have $OP*V_{\bar{p}}=V_{\bar{p}+1}H_{\bar{p}+1}$ with~:

  \[H_{\bar{p}+1}= \begin{pmatrix} h_{1,1} & h_{1,2} & \vdots & \vdots & h_{1j}& \vdots & h_{1,\bar{p}}\\ h_{2,1} & h_{2,2}& \vdots & \vdots & h_{2,j} & \vdots & h_{2,\bar{p}}\\0 & h_{3,2}& \vdots & \vdots & h_{3,j} & \vdots & h_{3,\bar{p}}\\ \vdots & \ddots & \ddots & \vdots &\vdots &\vdots &\vdots\\
  \vdots &  & \ddots & \ddots&\vdots&\vdots&\vdots\\ \vdots &  &  & \ddots&h_{j+1,j}&\vdots&h_{j+1,\bar{p}}\\ \vdots &  &  & &\ddots&\ddots&\vdots\\0 & \ldots & \ldots & \ldots& \ldots &0 & h_{\bar{p}+1,\bar{p}} \\ \end{pmatrix} 
.\]

Then let the upper Hessenberg matrix $H_{\bar{p}}$ be the matrix $H_{\bar{p}+1}$ without the $\bar{p}+1$ row. We have the equalities~:

\begin{equation}\label{stability}OP*V_{\bar{p}}=V_{\bar{p}}H_{\bar{p}}+h_{\bar{p}+1,\bar{p}}\mathbf{v}_{\bar{p}+1}\otimes \mathbf{e}_{\bar{p}}\end{equation}

This can also be written as~:

\[(OP-h_{\bar{p}+1,\bar{p}}\mathbf{v}_{\bar{p}+1}\otimes \mathbf{v}_{\bar{p}})V_{\bar{p}}=V_{\bar{p}}H_{\bar{p}}.\]

If we denote by $\mathbf{Z}_{\bar{p}}$ one eigenvector of the matrix $H_{\bar{p}}$, $V_{\bar{p}}\mathbf{Z}_{\bar{p}}$ is an eigenvector of the operator $OP$ and we have the following approximation~:
\begin{small}
\[ \Vert OP*V_{\bar{p}}\mathbf{Z}_{\bar{p}}-\lambda V_{\bar{p}} \mathbf{Z}_{\bar{p}}\Vert =\Vert (OP*V_{\bar{p}}-V_{\bar{p}}H_{\bar{p}} ) \mathbf{Z}_{\bar{p}} \Vert \leq\vert h_{\bar{p}+1,\bar{p}}\vert \Vert \mathbf{v}_{\bar{p}+1}\Vert_{\infty}\vert   z_{\bar{p}}\vert  \]
\end{small}

Let $ \bar{tol}$ be a fixed real positive quantity. It is considered that eigenvalue is acceptable if
\begin{equation} \label{Conv} \vert h_{\bar{p}+1,\bar{p}}\vert \Vert \mathbf{v}_{\bar{p}+1}\Vert_{\infty}\vert  z_{\bar{p}}\vert \leq \bar{tol}.\end{equation}

Afterwards the Implicit  Restarted Arnoldi Algorithm also called IRAM can be written as follows \cite{IRAMS}~:
\newline
Start: Build a length $\bar{p}$ Arnoldi factorization $OP*V_{\bar{p}}=V_{\bar{p}}H_{\bar{p}}+h_{\bar{p}+1,\bar{p}}\mathbf{v}_{\bar{p}+1}\otimes \mathbf{e}_{\bar{p}}$ with a starting vector ${\bf v}_1.$ 
\begin{enumerate}
\item Compute the eigenvalues $\{\lambda_j : j = 1,2,\ldots,\bar{k}\}$ by using classical QR algorithms \cite{AlgoQRVP} combined with Givens  rotations \cite{Givens,GivensC} applied to the matrix $H_{\bar{p}}$.
 \item If the first $\bar{k}$ eigenvalues verify the criterion \eqref{Conv}, then stop the algorithm.
\item Else Perform $\bar{p}-\bar{k}=m$ steps of the QR algorithm on the matrix $H_{\bar{p}}$ with the unwanted eigenvalues $\{ \lambda_j : j = \bar{k}+1, \bar{k}+2,\ldots,m \} $ as shifts \cite{AlgoQRVPShifted} to obtain $H_{\bar{p}}  Q_{\bar{p}} =  Q_{\bar{p}}  H_{\bar{p}}^+.$
\item Restart: Post multiply the length $\bar{p}$ Arnoldi factorization with the matrix $Q_{\bar{k}} $ consisting of the leading $\bar{k}$ columns of $Q_{\bar{p}} $ to obtain the length $\bar{k}$ Arnoldi factorization $ OP*V_{\bar{p}}  Q_{\bar{k}} =  V_{\bar{p}}  Q_{\bar{k}} H_{\bar{k}}^{+} + {\bf f}_{\bar{k}}^{+}\otimes {\bf e}_{\bar{k}}^T,$ where ${\bar{k}}^{+}$ is the leading principal sub-matrix of order $\bar{k}$ for $ H_{\bar{p}}^+.$ Set $ V_{\bar{k}} \leftarrow  V_{\bar{p}}  Q_{\bar{k}}.$
 \item Extend the length $\bar{k}$ Arnoldi factorization to a length $\bar{p}$ factorization and return to 1.
  \end{enumerate}

\subsection{Solution set up}
Let us make a quick summary about the use of the solver.
The input parameters are:
\vskip 0.3truecm
\begin{tabular}{|c|c|}
\hline
& \\
Two sparse and Hermitian &  $\tilde{A}$ and $\tilde{M}$  \\
finite element matrices. &  \\
& \\
\hline
& \\
Phase shift between   & $\phi=\mathbf{k}_i.\mathbf{T}$ \\
master slave faces & \\
& \\
\hline
& \\
Number of wanted eigenvalues & $\bar{k}$ \\
(Also called Ritz values) & \\
& \\
\hline
& \\
Number of Arnoldi vectors & $\bar{p}$ \\
& \\
\hline
& \\
Tolerance criteria  & $\bar{tol}$ \\

& \\
\hline
& \\
\textit{Shift} defined by the relation \eqref{defshift}, & $\sigma $ \\
near to the first solutions& \\
& \\
\hline
\end{tabular}
\vskip 0.3truecm
A good choice of the initial \textit{shift} is given by~:
\begin{equation}\label{shift}\sigma = 3\pi^2\min(\frac{1}{L^2},\frac{1}{l^2},\frac{1}{h^2})\sum_{V_i \in V_{tot}}\frac{1}{\epsilon_{r_{i}}  \mu_{r_{i}}}\frac{V_i}{V_{tot}} \end{equation}
Where $L$ correspond to the length, $l$ the width and $h$ to the height of the domain. $\epsilon_{r_{i}}, \mu_{r_{i}}$ are the relative permittivity and permeability of the different sub-domains and $V_i$ is their respective volume. 
This choice is motivated by the fact that the value of the first eigenmodes  in an homogeneous rectangular cavity filled with a dielectric $(\epsilon_r,\mu_r)$ is equal to \cite{Combes}~:
\[(k_0^2)_{mnp}= \frac{\pi^2}{\epsilon_r\mu_r}(\frac{p^2}{L^2}+\frac{m^2}{l^2}+\frac{n^2}{h^2}).\]
The value of the \textit{shift} $\sigma$  approximately corresponds to the first analytical eigenmode in a rectangular metallic cavity of the same size. The initial value given by the relation \eqref{shift}  is only an estimation is used as a point of departure for the solver. This is the reason we let the user the possibility to change this initial value. For example, when the periodic cell contains several resonant elements, the \textit{shift} can be adjusted to the wave number proportional with the wavelength corresponding to the size of one of these elements. Moreover, it is easy to understand that the parameter $\sigma$ depends on the number of  wanted eigenvalues $\bar{k}$. More precisely, if we increase the value of $\bar{k}$, we can choose $\sigma$ further from  to the first wanted eigenvalues.
Moreover, since the parameters are linked-together they do not all need to be filled. Indeed the numerical accuracy of the result will depend on the convergence criteria  $\bar{tol} $, the number of Arnoldi vectors $\bar{p}$ and the number of wanted eigenvalues $\bar{k}$. Due to the structure of the algorithm, we can easily understand that $\bar{p}$ must be greater than $\bar{k}$ and $\bar{p}$ must be large enough to avoid too many restarts of the IRAM \cite{IRAMS} algorithm. This is the reason we choose to link these two parameters as follows~: 
\begin{centered}
\includegraphics[width=0.6\linewidth]{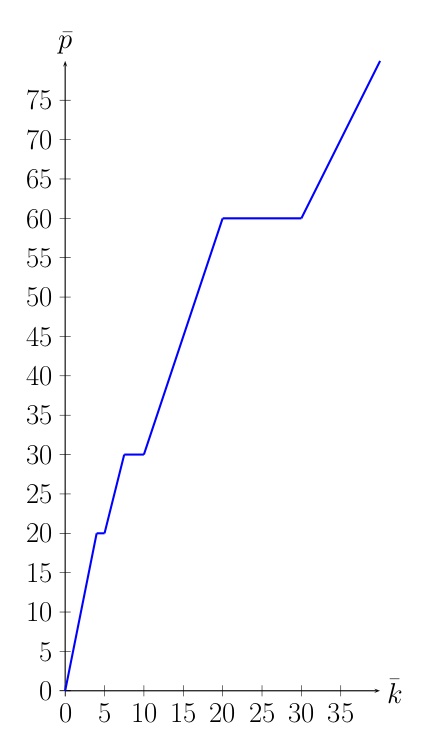}\\
\CPTFIG Relationships between $\bar{k}$ and $\bar{p}$.
\end{centered}
\CPTIDD{0} the number of Arnoldi vector $\bar{p}$ should be minimal to optimize memory usage, but if $\bar{p}$ is chosen close to $\bar {k}$, additional reboots  of the algorithm IRAM \cite{IRAMS} will be necessary for the first $\bar{k}$ eigenvalues to satisfy the criterion \eqref{Conv}. The most optimal choice is the one that will minimize both the number of restarts and the number of Arnoldi vectors $\bar{p}$. In the problems we are addressing, we have taken a pragmatic decision to minimize the time of resolution and to propose an automatic computation.

\begin{rem}
For each eigenmode solution $(k_0^2, \mathbf{X})$, we introduce the residue $\epsilon$ which will allow to eliminate null eigenvalues according to the criterion~: 
\begin{equation} \label{res} \begin{array}{c} \mathrm{if \ } \epsilon = \frac{\Vert A\mathbf{X}-k_0^2 M\mathbf{X}\Vert}{\vert k_0^2 \vert} \leq \bar{tol}\\
 \mathrm{ \ then \ } (k_0^2, \mathbf{X}) \mathrm{ \ is  \ an \ acceptable \ solution \ }.\end{array}\end{equation}
\end{rem}

It is well known that the number of unknowns will influence the physical accuracy of the result. Moreover, some part of the mesh domain has to contain more edges than others in order to compute values that correspond to the physical solutions. In order to know exactly which part we have to mesh thinner we can iterate on the mesh and compare the local variations of the solutions as HFFS do \cite{HFFSHELP1}. By doing this process we can observe than the thinner meshes have to be located in areas with highly diffracting materials or areas "narrow" where geometry varies abruptly. But this process is very time expensive, so we just need to get in a first step the optimized mesh.

\subsection{Sort algorithm}\label{Sort}

We present here an algorithm used to remove the zero solutions by changing the value of the \textit{shift} and using the criterion \eqref{res} during the calculation~: 
\begin{itemize}

\item Thanks to the Arnoldi algorithm, we compute the eigenvalues of the operator $(A-\sigma M)^{-1}M$ where the initial \textit{shift} $\sigma$ is defined by the relation \eqref{shift}.
\item We will obtain  $\bar{k}$ Ritz values for each given value of  $\mathbf{k}_i$.

\item $n_s$ of these solutions fulfil the convergence criteria \eqref{res}. These $n_s$ solutions are stored in the array called $Tab[]$.

 \item If $n_s < 2*\bar{k}/3$ , the value of the shift $\sigma$  increased , the solver is restarted and the values are not retained.
 For a fixed $\bar{k}$ we change the value of $\sigma$ (called $\sigma_{new}$) compared with the old value $\sigma_{old}$~:
\[\sigma_{new}=\frac{1}{4}\max(Tab[])+\frac{3}{4}\sigma_{old} \]
 
 \item Else If $n_s=\bar{k}$  the value of the shift $\sigma $  decreased , the solver is restarted and the values are not retained.
 For a fixed $\bar{k}$ we change the value of $\sigma$ (called $\sigma_{new}$) compared with the old value $\sigma_{old}$~:
 \[\sigma_{new}=\frac{1}{4}\min(Tab[])+\frac{3}{4}\sigma_{old} \]
 
  \item Else the values are retained and we change the value of $\sigma$ (called $\sigma_{new}$) compared with the old value $\sigma_{old}$:
\[\sigma_{new}= \frac{n_s-1}{n_s}\sigma_{old}+\frac{\overline{Tab[]}}{n_s},\] 
 where $\overline{Tab[]}$ means the average value of $Tab[]$.
 
\end{itemize}

Finally, for each value of $\mathbf{k}_i$ we will print all  eigenvalues $(k_0^2, \mathbf{X})$ that fulfil the convergence criteria \eqref{res} in a band diagram. We can also print the field created for each Ritz value.
If $\sigma$ is chosen too large, all solutions will verify the criteria \eqref{res}, ie $n_s$ is equal to $\bar {k}$ and the algorithm is restarted until  $n_s <\bar{k} $.
Thus the choice of the initial value of \textit{shift} $\sigma $ will play an important role in the computation time.

\section{Results}\label{RESULT}
\subsection{ Metallic Cubes}\label{Cubes}
 We consider perfectly conducting cubes left in air. The cell size is the original cube height $Dx = Dy = Dz = 1 m$, and the length of the side of the cube $w = 0.5m$. The characteristics $ \epsilon, \mu $ are those of the air  $ \epsilon_1 = \mu_1 = 1 $.  The value of the \textit{shift} $\sigma $ determined by \eqref{shift} is equal to 5 rad/m.

\begin{centered}
\includegraphics[width=0.5\linewidth]{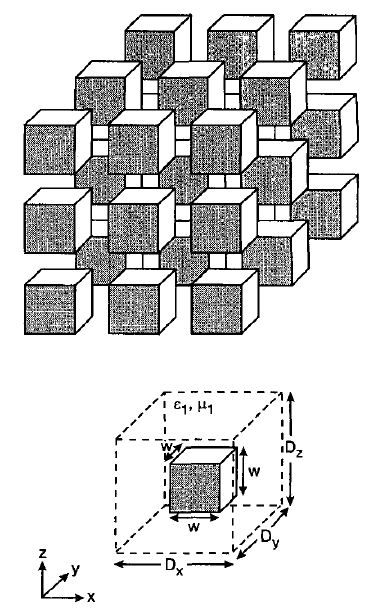} \\
\CPTFIG  Metallic Cubes in the air \cite{Periodique3D}.
\end{centered}
\vskip 0.2truecm

For our simulations, we  first vary the phase shift in the $Ox$ direction from $ 0 $ to $ \pi $ and the phase shifts in the other directions are set to zero. Then, we fix the phase shift in the $Ox$ direction to $ \pi $, that in the  $Oz$ direction to 0 and that in the $Oy$ direction varies from 0 to $ \pi $ (range $ [\pi, 2 \pi] $ in the diagram). Then, the phase shifts $Ox$ and $Oy$ are fixed at $ \pi $ and the $Oz$ phase shift varies from 0 to  $ \pi $ (range $ [2 \pi , 3 \pi] $ in the diagram). The results are expressed in radians per metre and the phase is given in radians. On the metallic parts, the tangent field is null. 
\begin{centered}
\includegraphics[width=1\linewidth]{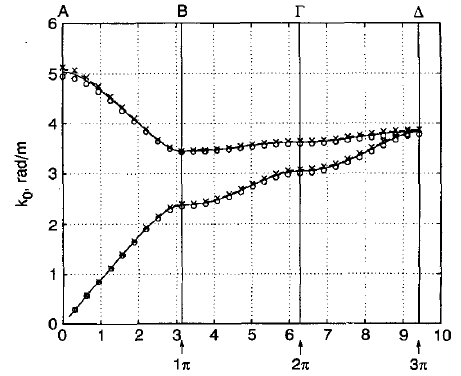} \\
\CPTFIG Results from the article\cite{Periodique3D}
\end{centered}
\vskip 0.2truecm
\begin{centered}
\includegraphics[width=1\linewidth ]{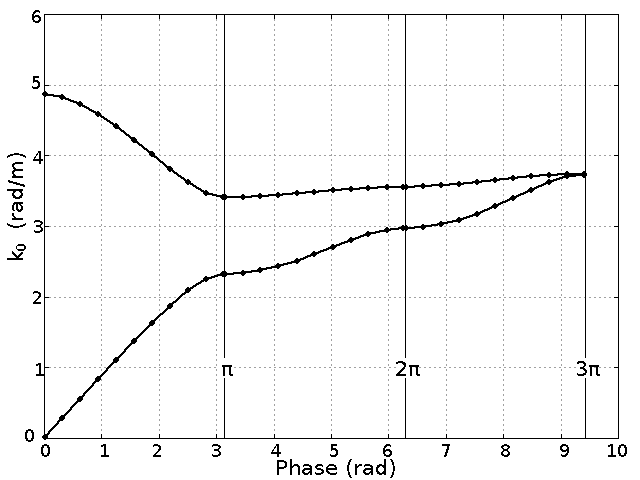} \\
\CPTFIG  Results from our solver.
\end{centered}
 In \CPTIDD{-1} the method used in the Article \cite{Periodique3D} is also a finite element method, but the algorithm used to compute the eigenmodes is not specified.  In \CPTIDD{-1} the number of unknowns is equal to 10344 and  in \CPTIDD{0} 9599. This explains the very small difference between the two diagrams. To conclude this example, we can say that  the results are validated  and that it is possible to extend the method to non-canonical geometries.

\begin{rem} The FDTD\cite{Disper4,FDTDCFL} method can also treat this kind of problem but if we change metallic cubes to 3D complex structures (cylinders or pyramids for example); since the meshed area is necessary cubic and the geometry must be meshed uniform way and small enough to ensure the stability of the numerical scheme \cite{FDTDCFL}, computation time and memory space required can drastically increased. Our method accepts all of the kinds of meshes, unless the master slave relations are respected on the master slave faces. Thus we have less restrictions on the geometry. 
 \end{rem}
 
\subsection{Infinite dielectric rods}
\paragraph{Basic cell}
Let us consider a square network of infinite dielectric rods in the $Oz$ direction. The entries on the axes are the components of the plane wave excitation.
\begin{centered}
\includegraphics[width=0.6\linewidth]{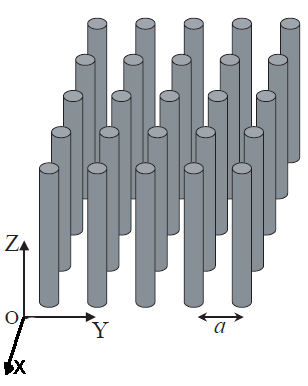} \\
\CPTFIG Square network of infinite dielectric rods \\
a=7mm, radius=2mm, $\epsilon_r=9.4$. \\
\end{centered}
\begin{centered}
\includegraphics[width=0.4\linewidth]{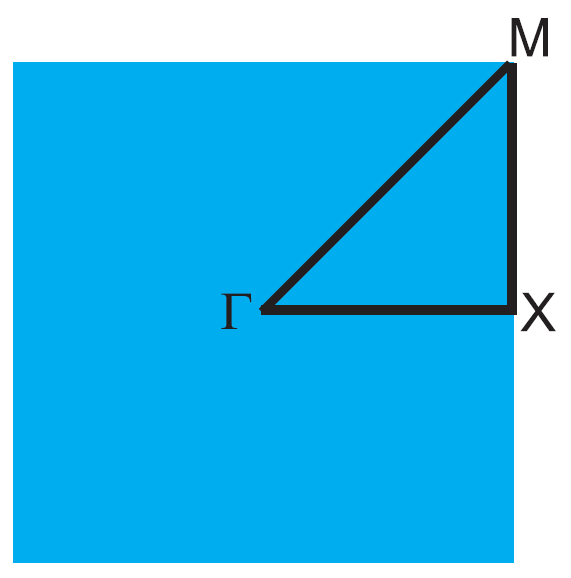}\\
\CPTFIG We vary the phase shift in the basic cell.\\
The phase shift in the Oz direction is set to 0.\\
The shift $\sigma$ defined by \eqref{shift} is equal to 21 Giga Hertz.\\
 \end{centered}
 \vskip 0.2true cm
The geometry is invariant in the OZ direction, then we set $k_{iz}=0$ and the height of the meshed area is chosen thinner than the length and the width~:
 \begin{centered}
\includegraphics[width=1\linewidth]{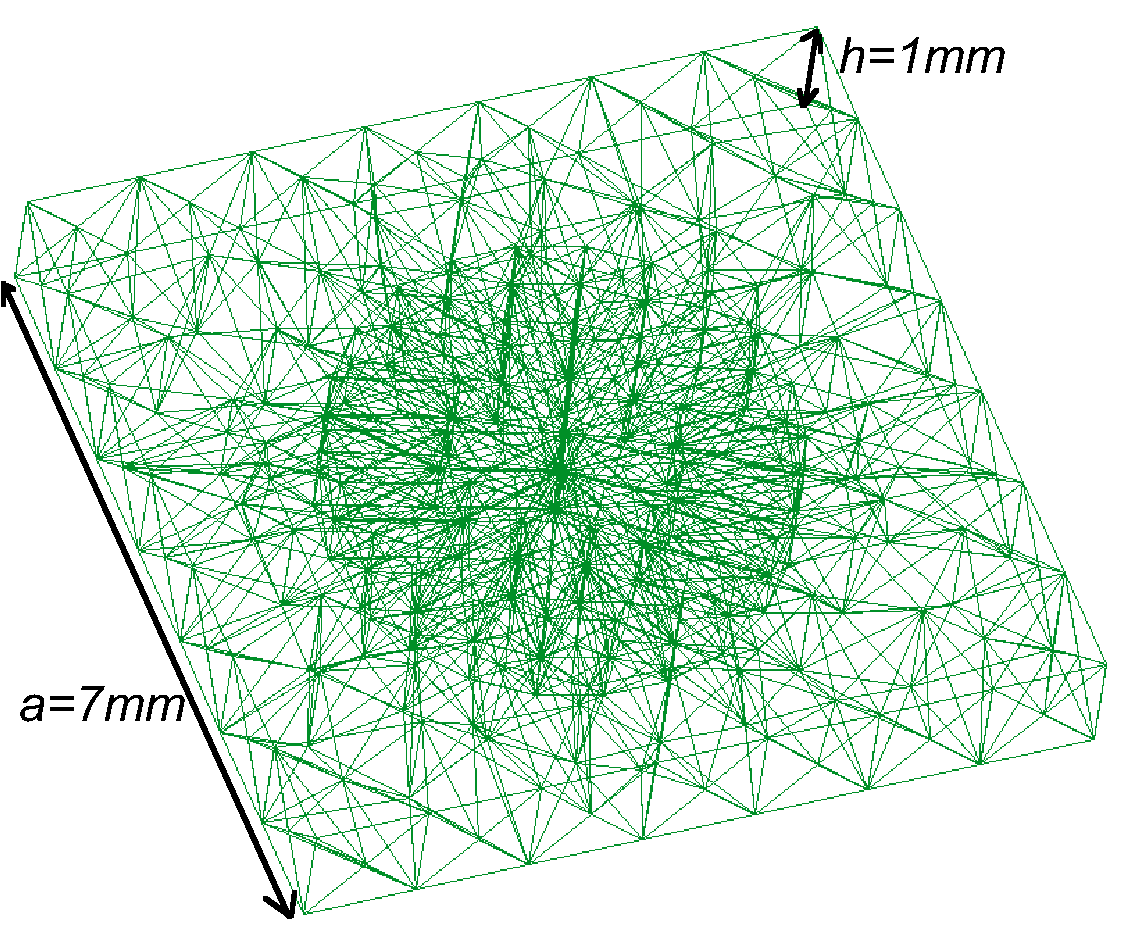}\\
\CPTFIG Meshed area of the basic cell  \\
 with a thickness=1mm.\\
 \end{centered}
 
 \begin{centered}
\includegraphics[width=1\linewidth]{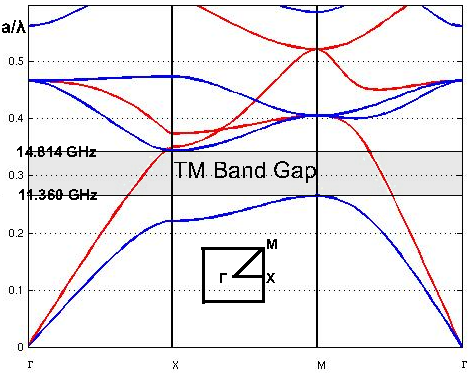}\\
\CPTFIG Results of the Plane wave method\cite{Cristphoto2}.\\
\textcolor{blue}{TMZ  modes}, \textcolor{red}{TEZ modes}.
 \end{centered}
 \vskip 0.2true cm
  \begin{centered}
\includegraphics[width=1\linewidth]{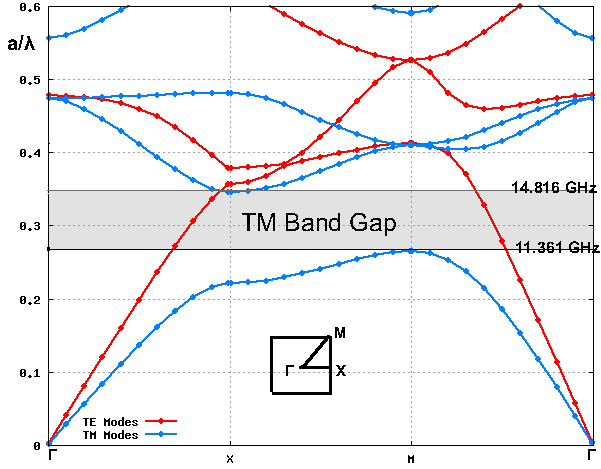}\\
\CPTFIG Results from our solver.\\
\textcolor{blue}{TMZ  modes}, \textcolor{red}{TEZ modes}.
\end{centered}
%\label{fig-12}
%\end{figure}
As we can see, the plane wave diagram can be accurately reprinted with our method. To treat this case, it is also possible to use the finite element method for two-dimensional doubly periodic structures, as it is developed in \cite{Periodique3D}. Due to the geometry, the solution can easily be computed with the plane wave method using Bessels functions \cite{Cristophoto}.

\begin{rem}
We can deal with this problem by using HFFS, but in the latest version the solver will not give the solutions under the minimum chosen frequency \cite{HFFSHELP2}. This can be problematic if the first solutions are very near to zero, as we can see in this case because if we set this frequency to a  small value the results will be corrupted \cite{HFFSHELP1}. In addition, if  this frequency is set to a value that is too large, we will not get the first eigenmodes.
\end{rem}

\subsection{Super-cells}
 Periodic patterns which will have more than one basic cell are called super cells.
 The basic cell is now defined by a=14mm, radius=0.18a and  $\epsilon_r = 11.56  $.
 We want to study the propagation in the $Ox,Oy$ plane (\CPTIDD{2}, \CPTIDD{6}), so the $k_{iz}$ component is fixed to zero and the thickness of the cells is equal to 1mm.  We vary the phase shift $k_{iy}=\phi_y/dy$ for multiple values of $k_{ix}=\phi_x/dx$. In other words, we compute the projected band diagram \cite{PeriodicBook} in the $Oy$ direction. 

\begin{centered}
\includegraphics[width=1\linewidth]{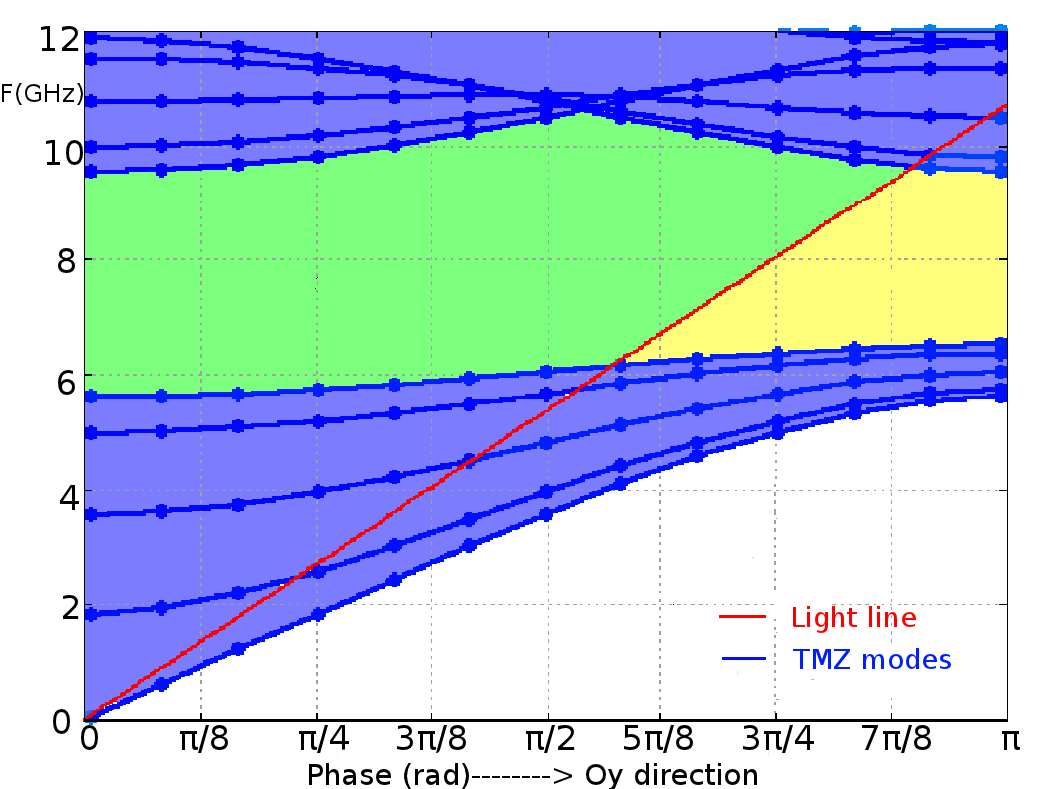} \\
\CPTFIG Projected ban diagram for  $\epsilon_r = 11.56$ TMZ solutions.\\
$\phi_x\in[0,\pi,\pi/2,3*\pi/4,\pi/4] \mathrm{ \ and \ } \phi_y\in[0,\pi/8,\pi/4,\ldots,\pi]$.\\
 $\sigma$ = 12.9 \eqref{shift}.
\setcounter{moncompteur}{\thecpfig}
\end{centered}

In \CPTIDD{0}, the straight red line represents the light line defined by the equation $\omega = c |\mathbf{k}_{||}|$.
We call guided modes that are able to propagate in the air but not into the meta-material. These appear when we introduce a linear defect into the material; the existence of these modes corresponds to the green zone. Modes that are evanescent in the air and in the meta-material are called surfaces modes. These correspond to the yellow zone and appear when we change the size of the last elements of the periodical structures \cite{PeriodicBook}. To trap these modes, we establish the band diagrams of  super-cells \cite{PeriodicBook}. The $Oy$ direction corresponds to the direction of the defect introduced in the super-cells.
\paragraph{Linear defect}
\begin{centered}
\includegraphics[width=1\linewidth]{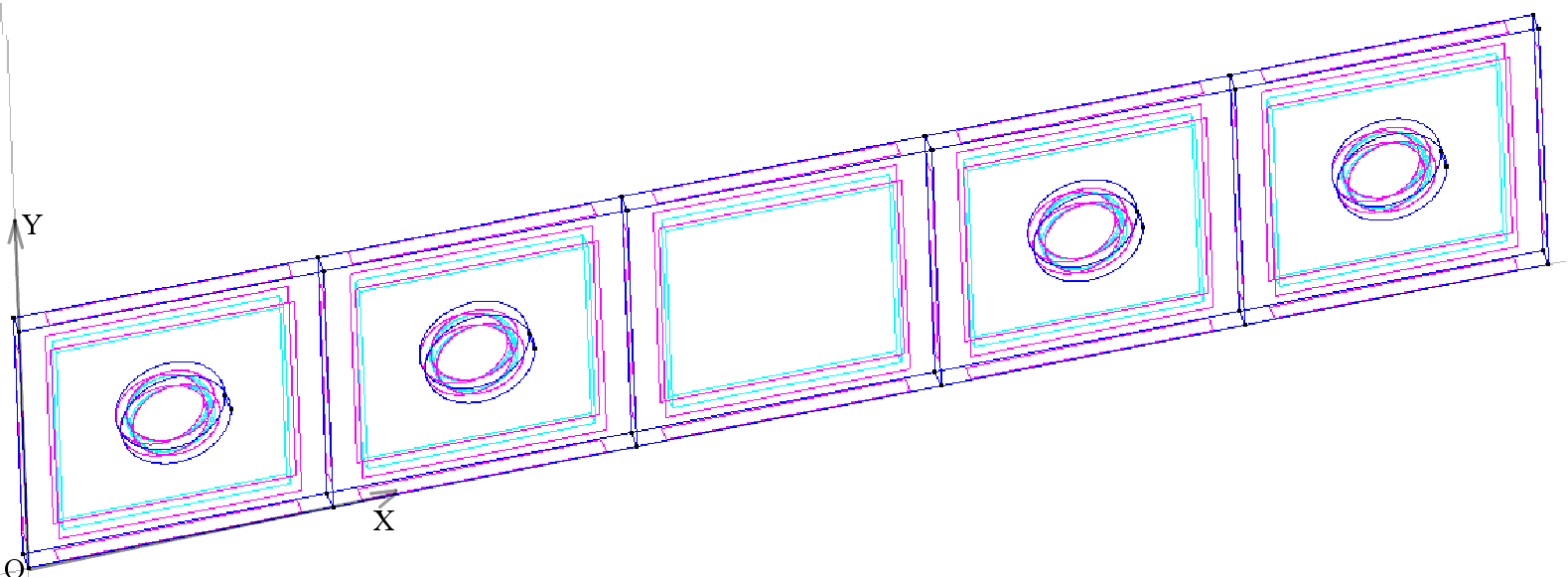}\\
\CPTFIG Linear defect introduced into the periodic structure.\\
\setcounter{compp}{\thecpfig}
\end{centered}
\vskip 0.2truecm
\begin{rem}
The shift $\sigma$ defined by \eqref{shift} will be too small because the dimensions of the super cell are much larger than the dimensions of the basic cell. Then, we choose a shift included in the band gap that we had already found by studying the basic cell. 
\end{rem}
The corresponding band-diagram for the super-cell \CPTIDD{0} is~:
\begin{centered}
\includegraphics[width=1\linewidth]{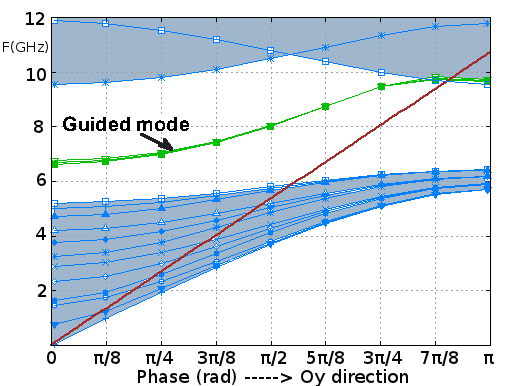}\\
\CPTFIG Projected band diagram  for super cellule with a linear defect.\\
$\phi_x\in[0,\pi/2,\pi,3*\pi/4] \mathrm{ \ and \ } \phi_y\in[0,\pi/8,\pi/4,\ldots,\pi]$,$\epsilon_r=11.56$.\\
 k=15, $\epsilon=10^{-4}$ and $\sigma=7.5 \mathrm{ \ GigaHz }$.
\end{centered}

In \CPTIDD{0} each point $(\phi_x,\phi_y)$ will give 12 over 15 eigenvalues which will verify the criteria \eqref{res}.

Here we can observe the values of the $Re(E_z)$ and $\vert E_z \vert $ guided fields on the plane z=0.0005. The frequency of the mode is 8.58 Giga Hertz~:

\begin{centered}
\includegraphics[width=0.4\linewidth]{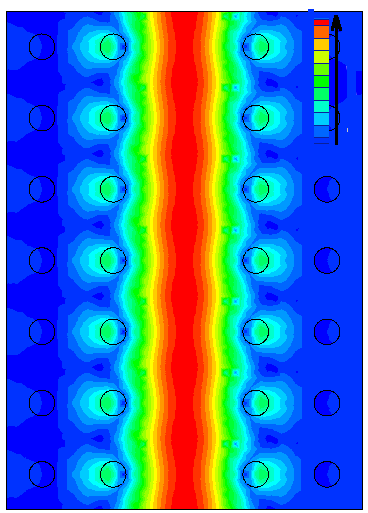}\\
\CPTFIG \\
  $|E_z|$ for $k_{ix}=\pi/dx$ , $k_{iy}=0.6\pi/dy$.\\

\end{centered}

\begin{centered}
\includegraphics[width=0.4\linewidth]{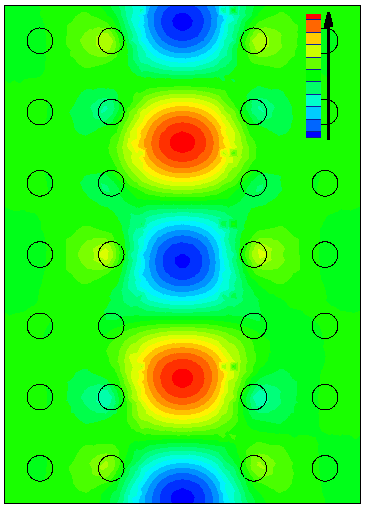}\\
\CPTFIG \\
 $Re(E_z)$ for $k_{ix}=\pi/dx$ , $k_{iy}=0.6\pi/dy$.
\end{centered}

\subsubsection{Surface defect}
We now will introduce a defect by modifying the last row of the super-cell~:
\begin{centered}
\includegraphics[width=1\linewidth]{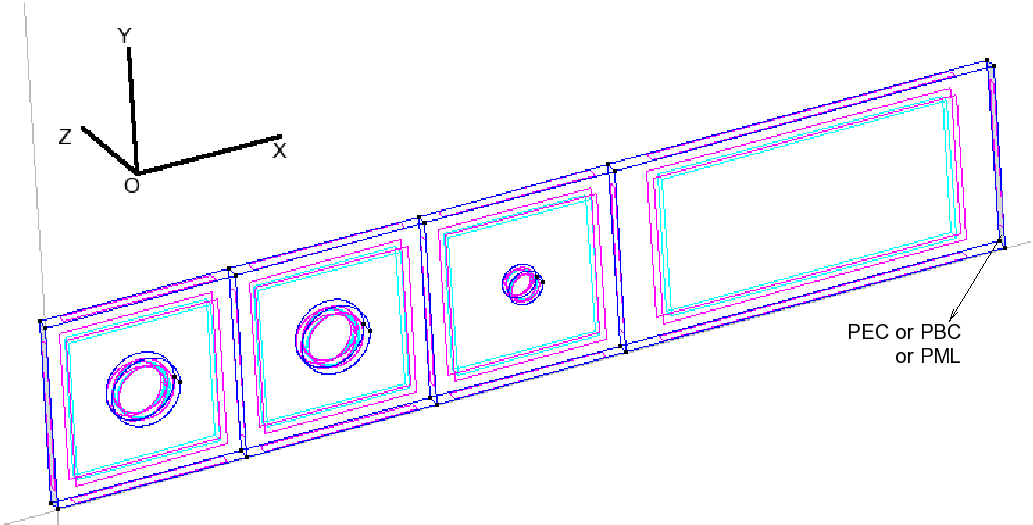}\\
\CPTFIG Modification of the last rods to make surface modes appear.\\
The radius of the surface rods is equal to 0.009a.\\
\setcounter{comppp}{\thecpfig}
\end{centered}
\vskip 0.2truecm
\begin{rem}
 In \cite{PeriodicBook, FloquetBloch} the plane wave method is used to characterize surface modes  when the network  is defined by square rods or circular rods. This emphasizes the fact that the last rod must be different from the rods of the network, in the example the last rod is cut down its middle.
\end{rem}

The corresponding band-diagram for the \CPTIDD{0} is~:

\begin{centered}
\includegraphics[width=1\linewidth]{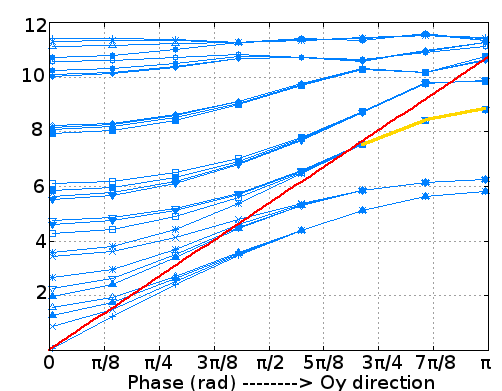}\\
\CPTFIG Projected band diagram  for a super cell with modification of the surface rods in the Oy direction.\\
$\phi_x\in[0,\pi,\pi/2,3*\pi/4] \mathrm{ \ and \ } \phi_y\in[0,\pi/8,\pi/4,\ldots,\pi]$.\\
\end{centered}
\CPTIDD{0} The surface modes are represented in yellow.
We can now observe the values of the $Re(E_z)$ and $\vert E_z \vert$ field. The frequency of the mode is 8.51 Giga-Hertz (images taken with the software Tecplot)~:

\begin{centered}
\includegraphics[width=0.4\linewidth]{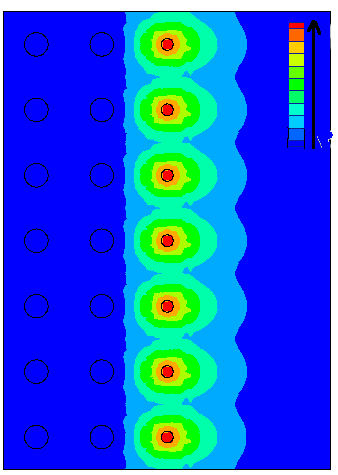}\\
\CPTFIG \\
 $|E_z|$ for $k_{ix}=0$, $k_{iy}=7\pi/8dy$.\\
 
\end{centered}

\begin{centered}
\includegraphics[width=0.4\linewidth]{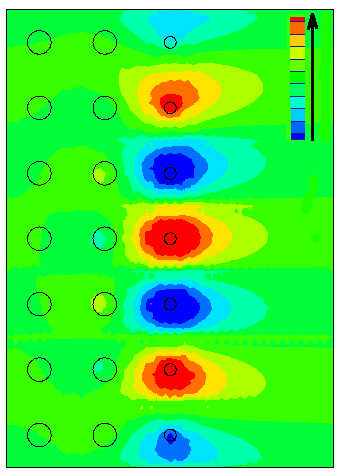}\\
\CPTFIG \\
  $Re(E_z)$ for $k_{ix}=0$ , $k_{iy}=7\pi/8dy$.\\
 
\end{centered}

\begin{rem}
 By modulating the value of the shift $\sigma$, we can directly seek solutions that are located in the red and yellow part and then there is no need to consider a lot of wanted eigenvalues $\bar{k}$. For example, if we only seek surface mode solutions we can take  $\bar{k}=2$ and $(\sigma,\phi_y)$ located in the yellow zone in \CPTID.
\end{rem}

\section{Summary of the computational performances}\label{CompPer}
We now make a summary for each case of the computational performances treated on our Intel (R) Xeon 64 bits 2.80 GHZ machine. We specify the CPU time, the total number of edges in the mesh file, the RAM used during the process and the number of points in the band-diagram.The number of wanted eigenvalues $\bar{k}$ is always equal to 15 and the tolerance criteria $\bar{tol}$ is always equal to $10^{-7}$.
\vskip 0.2true cm
\begin{tabular}{|c|c|c|c|c|}
\hline
& & & &\\
& Edges & RAM  & Time &  Points \\
& & & & \\
\hline
& & & &\\
Metallic &9599&152.7 Mb & 4m33s &33 \\ 
cubes & & & &\\
& & & & \\
\hline
& & & &\\
Dielectric&7035&49 Mb &46s &33 \\ 
rods & & & &\\
& & & &\\
\hline
& & & &\\
Linear &12683 &72 Mb & 94s &36 \\ 
defect &&&&\\
& & & &\\
\hline
& & & &\\
Surface &10471&60 Mb &67s &36 \\ 
defect &&&&\\
& & & &\\
\hline
\end{tabular}

\section{Conclusion and perspective}
We have presented the theoretical aspect of the eigenmode solver for periodic non-dispersive structures first. Then, we have presented the way to combine different libraries, such as ARPACK and PARDISO, to optimize the Arnoldi algorithm and finally, we have presented the results on the original periodic structures. We are now trying to study a dispersive structure by adapting a non-linear plane wave technique  \cite{Disper2} to our model. 
\section*{Acknowledgment}
The authors would like to thank Stephane Varault from ONERA,
for providing comparative results using the plane wave method for the infinite dielectric rods.

\bibliographystyle{model1-num-names}

\bibliography{Biblio}\vskip 0.2truecm

\end{document}